\documentclass[unnumsec,webpdf,contemporary,large]{oup-authoring-template}%

\usepackage{subfig}
\usepackage{framed}

\usepackage{todonotes}

\graphicspath{{Fig/}}

\theoremstyle{thmstyleone}%
\theoremstyle{thmstyletwo}%
\theoremstyle{thmstylethree}%

\begin{document}

\journaltitle{OUP Synthetic Biology}
\DOI{DOI HERE}
\copyrightyear{2022}
\pubyear{2022}
\access{Advance Access Publication Date: Day Month Year}
\appnotes{Paper}

\firstpage{1}


\title[Functional Synthetic Biology]{Functional Synthetic Biology}

\author[1]{Ibrahim Aldulijan\ORCID{0000-0001-7332-619X}}
\author[2,$\ast$]{Jacob Beal\ORCID{0000-0002-1663-5102}}
\author[3]{Sonja Billerbeck}
\author[4]{Jeff Bouffard\ORCID{0000-0001-9121-8416}}
\author[5]{Gaël Chambonnier\ORCID{0000-0002-5606-2089}}
\author[6]{Nikolaos Delkis\ORCID{0000-0003-2310-4040}}	
\author[7]{Isaac Guerreiro}
\author[8]{Martin Holub}
\author[9]{Paul Ross}
\author[7]{Vinoo Selvarajah}
\author[10]{Noah Sprent\ORCID{0000-0001-9129-9236}}
\author[11]{Gonzalo Vidal\ORCID{0000-0003-3543-520X}}
\author[12]{Alejandro Vignoni\ORCID{0000-0001-9977-7132}}

\authormark{Aldulijan et al.}

\address[1]{\orgdiv{Systems Engineering}, \orgname{Stevens Institute of Technology}, \orgaddress{\street{1 Castle Point Terrace, Hoboken}, \postcode{07030}, \state{NJ}, \country{USA}}}
\address[2]{\orgdiv{Intelligent Software \& Systems}, \orgname{Raytheon BBN Technologies}, \orgaddress{\street{10 Moulton Street, Cambridge}, \postcode{02138}, \state{MA}, \country{USA}}}
\address[3]{\orgdiv{Molecular Microbiology, Groningen Biomolecular Sciences and Biotechnology Institute}, \orgname{University of Groningen}, \orgaddress{\street{Nijenborgh 7}, \postcode{9747 AG}, \state{Groningen}, \country{The Netherlands}}}
\address[4]{\orgdiv{Centre for Applied Synthetic Biology, and Department of Biology}, \orgname{Concordia University}, \orgaddress{\street{7141 Sherbrooke Street West, Montréal}, \postcode{H4B 1R6}, \state{Québec}, \country{Canada}}}
\address[5]{\orgdiv{Department of Biological Engineering}, \orgname{Massachusetts Institute of Technology}, \orgaddress{\street{Cambridge}, \postcode{02139}, \state{Massachusetts}, \country{USA}}}
\address[6]{\orgdiv{Plant and Environmental Biotechnology Laboratory, Department of Biochemistry and Biotechnology}, \orgname{University of Thessaly}, \orgaddress{\street{Viopolis, Mezourlo, Larissa}, \postcode{41500}, \country{Greece}}}
\address[7]{\orgname{iGEM Foundation}, \orgaddress{\street{45 Prospect Street, Cambridge}, 
\postcode{02139}, \state{MA}, \country{USA}}}
\address[8]{\orgname{Delft University of Technology}, \orgaddress{\street{Van
der Maasweg 9}, \postcode{2629 HZ}, \country{The Netherlands}}}
\address[9]{\orgname{BioStrat Marketing}, \orgaddress{\street{9965 Harbour Lake Circle, Boynton Beach}, \state{FL}, \postcode{33437}, \country{USA}}}
\address[10]{\orgdiv{Department of Chemical Engineering}, \orgname{Imperial College London}, \orgaddress{\street{South Kensington Campus, Exhibition Road}, \postcode{SW7 2AZ}, \country{UK}}}
\address[11]{\orgdiv{Interdisciplinary Computing and Complex BioSystems (ICOS) research group, School of Computing}, \orgname{Newcastle University}, \orgaddress{\street{Devonshire Building, Devonshire Terrace}, \postcode{NE1 7RU}, \state{Newcastle Upon Tyne}, \country{UK}}}
\address[12]{\orgdiv{Synthetic Biology and Biosystems Control Lab, Instituto de Automatica e Informatica Industrial}, \orgname{Universitat Politecnica de Valencia}, \orgaddress{\street{Camino de Vera s/n}, \postcode{46022}, \state{Valencia}, \country{Spain}}}

\corresp[$\ast$]{Corresponding author. \href{email:jakebeal@ieee.org}{jakebeal@ieee.org}}

\received{Date}{0}{Year}
\revised{Date}{0}{Year}
\accepted{Date}{0}{Year}



\abstract{
Synthetic biologists have made great progress over the past decade in developing methods for modular assembly of genetic sequences and in engineering biological systems with a wide variety of functions in various contexts and organisms. 
However, current paradigms in the field entangle sequence and functionality in a manner that makes abstraction difficult, reduces engineering flexibility, and impairs predictability and design reuse.
Functional Synthetic Biology aims to overcome these impediments by focusing the design of biological systems on function, rather than on sequence.
This reorientation will decouple the engineering of biological devices from the specifics of how those devices are put to use, requiring both conceptual and organizational change, as well as supporting software tooling.
Realizing this vision of Functional Synthetic Biology will allow more flexibility in how devices are used, more opportunity for reuse of devices and data, improvements in predictability, and reductions in technical risk and cost.
}



\maketitle


\section{Introduction}

Over the past decade, synthetic biologists have made great strides in the engineering of biological systems.
One vision that has served as something of a roadmap is the model enunciated by Endy~\cite{Endy05}, a model comprising four levels of increasing abstraction: DNA, Parts, Devices, and Systems.
Consistent with this model, the field has developed a plethora of basic parts such as promoters, terminators, coding sequences, and functional RNAs, which can be combined into composite DNA sequences through a variety of assembly methods (e.g., ~\cite{shetty2011assembly,gibson2009enzymatic,Engler2008GoldenGate,Weber2011MoClo,kok2014rapid}) or low-cost nucleic acid synthesis~\cite{CarlsonCurve}.
At higher levels of abstraction, the field has produced families of biological devices with a variety of sensing, communication, or computational functions (e.g.,~\cite{endy-pnas-2012, kiani2014crispr, nielsen2016genetic, gander2017digital, weinberg2019high, chen2020genetic, sexton2020multiplexing, kiwimagi21quantitative}),
as well as methods for insulating devices from context (e.g.,~\cite{RiboJ12, mutalik2013BCDs, carr2017reducing}), and for characterizing and predicting their behavior (e.g.,~\cite{del2008modular, Salis09, beal2015replicon, davidsohn2015accurate, nielsen2016genetic, wang2019modeling, chen2020genetic, sexton2020multiplexing, kiwimagi21quantitative, beal2018quantification,beal2020robust,pine2016evaluation}).

Despite this progress, significant challenges remain in the practice of synthetic biology at a systems level.
Definitions for parts and devices are typically unavailable, incomplete, or inconsistent.
Likewise, little information is generally provided regarding interfaces, functionality, or host context, and such information is rarely available in a tool-friendly format.
This leads to significant difficulty in searching for appropriate parts or devices to use, in adapting parts and devices from their original context for use within a new project, and in predicting the behavior of even the most basic systems from information available about the components used to create them.
These difficulties in turn lead to significant time requirements and technical risk to achieve even modest engineering goals.

Under these conditions, engineering success can certainly be achieved, as illustrated by the many billions of dollars of industrial impact from synthetic biology, but it is slow and costly to do so.
On the other hand, if engineering could be made simple and predictable, even just for systems comprising small numbers of devices, it would radically lower cost, reduce barriers to access, increase democratization, and unleash a wave of innovative applications of synthetic biology by small organizations tackling local problems.

We contend that the primary barrier to achieving this vision is no longer biological, given the myriad advances that have been achieved.
Rather, we argue that the problem is one of knowledge synthesis and organization: given the complexity of biological systems, no practitioner can be expected to even be aware of all of the relevant parts, methods, and models, 
let alone have the detailed expertise in all of them to use and combine them effectively in practice.
How then can the advances and expertise dispersed across the synthetic biology community be marshaled in order to enable practitioners to effectively utilize them in their engineering projects?

We propose that in order to meet this challenge, synthetic biology must be reoriented from its current sequence-centric approach to instead center on function.
Specifically, a Functional Synthetic Biology approach focuses on:
\begin{itemize}
\item descriptions of behavior over descriptions of structure,
\item predictability and flexibility over optimization of function, and
\item risk reduction over novelty.
\end{itemize}
A focus on behavior means a biological component's structure (i.e., genetic sequence) should be able to be changed and improved without damaging the functionality of a system that includes it.
A focus on predictability means identifying classes of changes that are unlikely to damage functionality, and a focus on flexibility means valuing the breadth of such classes of changes when developing biological components.
A focus on risk reduction means recognizing that there are many failure modes that can impair the functionality of a biological system, and that there is value in capturing knowledge about failure modes (and how to avoid them) in the form of automation tools and machine-readable component specifications.

\begin{figure*}[t!]
\centering
\includegraphics[width=\textwidth]{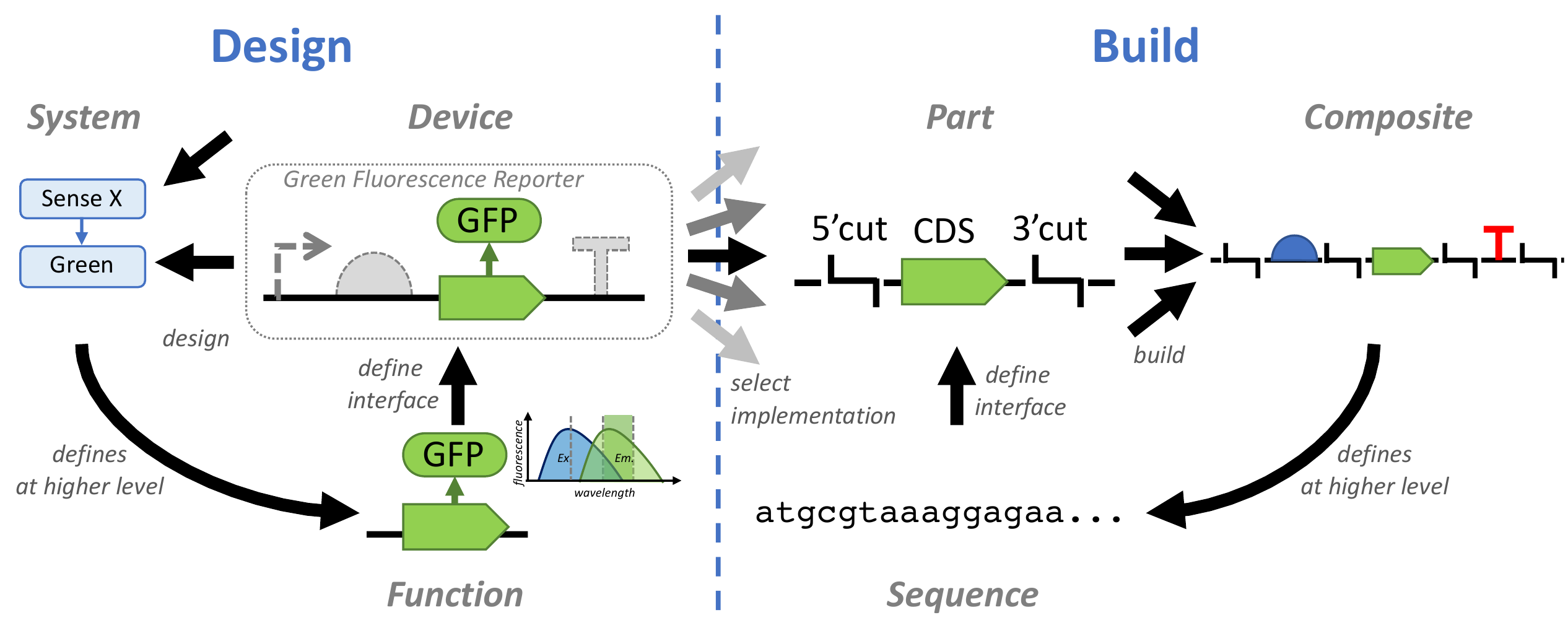}
\caption{Abstraction layers in a function-centric view: 
design focuses on biological functions, which are abstracted to produce devices by adding a description of an interface, predicted range of behavior, and operational context.
Devices may then be combined to produce a biological system, whose function in turn may be abstracted to create a new device at a higher level of abstraction.
Devices may have many different options for how they can be implemented with parts, 
where parts are sequences with a defined interface for combining them to build composite sequences, which also may in turn be abstracted into higher-level parts.}
\label{f:function}
\end{figure*}

Together, these approaches will allow the community of synthetic biologists to more effectively share successes and avoid failures.
Below, we will expand on each of these three key goals—describing behavior, predictability and flexibility, and risk reduction—and lay out a roadmap for achieving these goals over the course of the coming decade.

\section{Function-Centric Design Descriptions}

Current practices and representations in synthetic biology are mainly sequence-centric, meaning that the first-class objects of a design are sequences---typically DNA, though sometimes RNA or amino acids.
Information about function is then annotated as features or metadata used to describe the sequence.
Many design representations (e.g., FASTA, GenBank, GFF), reinforce this notion by providing no way of expressing functional information without reference to a fully-specified sequence.

In practice, however, synthetic biologists tend to think about designs more in terms of function.
Consider, for example, green fluorescent protein (GFP) as a reporter of transcriptional activity (left half of Figure~\ref{f:function}).
A typical user of GFP does not think about the actual sequences and would be unlikely even to recognize either the nucleic acid or amino acid sequences for GFP if they saw them.
Instead they are likely to think about a functional relationship between a coding sequence (CDS) that produces a GFP protein, which in turn will fluoresce with a predictable excitation and emission behavior.
This is a fully coherent notion of biological function, independent of sequence, and is the typical subject of discussions and diagrams regarding design.

Also associated with the notion of function is a concept of an interface to that function (e.g., embedding the GFP in a transcriptional unit whose activity is to be reported) and of the type of environments where the function is expected to behave as predicted (e.g., aerobic environments with relatively neutral pH across a broad range of cell types).
For Functional Synthetic Biology, then, we will define a device as a function that has been associated with a definition of an interface and a context for its operation.
This information is sufficient to combine a device together with others to implement a biological system (e.g., to sense some condition and report it via green fluorescence).
Finally, the function of that system may be itself described and possibly abstracted into a higher-level device by identifying its interface and context for operation.

To actually implement a device or system, of course, a specific sequence must be identified.
Typically there are many sequences that may suffice to implement any given device.
GFP, for example, at the time of this writing has 160 different amino acid sequences on FPbase~\cite{lambert2019fpbase}.
Reverse translation of any of these amino acid sequences further produces myriad distinct nucleic acid sequences, all of which encode the same amino acid sequence.
These alternatives are not identical, of course, and likely some will not be functional at all. Still, a great many are expected to be sufficiently good implementations of a Green Fluorescence Reporter device.

As with devices, we often need to combine sequences together with other sequences in order to form the composites that implement systems, whether through assembly protocols or by directly synthesizing the composite sequence.
The right half of Figure~\ref{f:function} shows this parallel abstraction hierarchy, in which a part is defined as a sequence that has been associated with a definition of a build interface, parts combine to form composites, and those composites may in turn be abstracted into higher-level parts.

\begin{figure*}
\centering
\includegraphics[width=\textwidth]{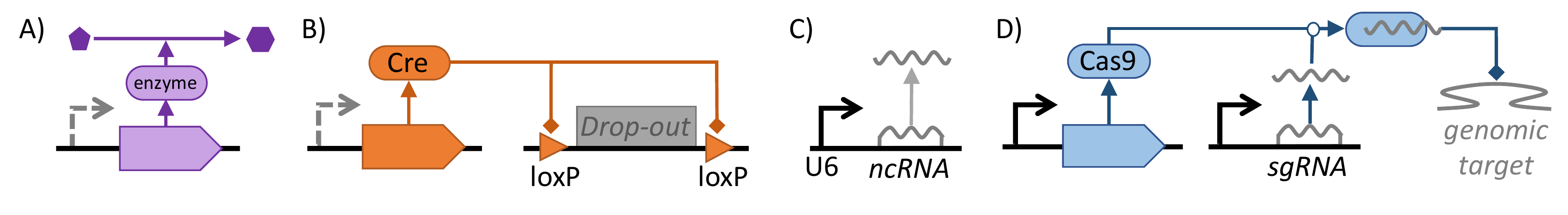}
\caption{Other examples of functional devices: 
A) an enzyme that catalyzes the transformation of one small molecule chemical into another,
B) excision of a drop-out sequence using Cre recombinase targeting loxP sites,
C) constitutive expression of non-coding RNA using a U6 promoter, and
D) CRISPR-based gene editing with constitutive expression of Cas9 and sgRNA.}
\label{f:devices}
\end{figure*}

This approach to defining functional devices can be applied to any well-described biological function: Figure~\ref{f:devices} shows other examples of functional devices, some of which would necessarily need to be implemented using multiple parts.

Indeed, this separated hierarchy is {\it de facto} what is already in informal use throughout much of the community. 
From whiteboards to journal articles, synthetic biologists typically communicate in terms of function. 
Sequences, meanwhile, are often selected arbitrarily or inherited from prior projects via a shared laboratory freezer, and in publications are typically relegated to supplementary information or even---problematically---entirely omitted~\cite{peccoud2011essential}.
Functional Synthetic Biology proposes that we recognize this distinction and the separation between sequence and function, then explore its consequences in order to improve our representations, tooling, engineering approaches, and collaboration strategies.

\section{Predictability and Flexibility}

The importance of predictability and flexibility as design requirements is heightened when a device's functional definition is divorced from its realization as a composite of discrete parts. 
Implementation of a function, whether at the level of a single device or a composite system, is a matter of identifying an appropriate part or sequence, and typically there are many that could potentially apply.

With sequence-centric engineering, characterization of a part has tended to ask questions of the form ``How does this part behave?''
Given a functional definition of a device, however, a fundamentally different question needs to be answered: 
``Is this part's behavior good enough to be an implementation for that device?''
This question cannot be answered without considering what ``good enough'' means in terms of the function of a device.

For example, a Green Fluorescence Reporter device turns transcriptional activity into a strong fluorescent signal.
How strong is strong enough? 
The answer is determined by the signal strength required to discriminate various levels of expression from background with a given class of instrument (e.g., plate reader or flow cytometer), which in turn determines the flexibility of the specifications for the device instantiation, i.e., what types and degrees of imperfections can be tolerated.
Device context must also be recorded in the specification, as any GFP coding sequence will fail to produce strong fluorescence if it is placed in the wrong operational context (e.g., in anaerobic conditions, with an incompatible 5' UTR, or in an incompatible host).
Useful device specifications thus require at least some predictions about device behavior. 

There is an inherent tension and interplay between predictability and flexibility. 
Any device can be made impossible to realize by making its specification require too much precision and/or too great an operational range, e.g., looking for a Green Fluorescence Reporter that always produces exactly the same number of molecules in wildly different cell types.
Likewise, a device can be rendered inoperable if the constraints placed on values are too lax (e.g., accepting a red fluorescent protein as a legitimate implementation of a Green Fluorescence Reporter) or if the operational range applied is too tightly constrained (e.g., predicting that the device will work only in the exact construct where it was characterized).
Success in designing and building engineered systems depends on finding a middle ground where practitioners are readily able both to find devices and parts to realize a system and also to predict the outcome that system will generate with a satisfactory degree of reliability and accuracy.

There is likewise a tension between these goals and the desire to obtain the best performance from a device.
For example, in selecting a part to realize a Green Fluorescence Reporter device, it may be desirable to select a less bright GFP variant that is better understood (thus more predictable), known to operate in a wider range of organisms, or available in a preferred assembly format.
Ultimately, this is a matter of trust and focus: the more that a practitioner can trust the predictability of the less interesting parts of their system (e.g., the Green Fluorescence Reporter), the more that they can focus on their primary goals (e.g., improvement of a novel metabolite sensor whose activity is being reported).

Navigating the relationship between flexibility, predictability, and optimization of function is in general a challenging and unresolved problem in all engineering fields, and particularly so for biology.
Nevertheless, there are a number of bioengineering tools, such as GFP, that are in common use precisely because they are reasonably effective at producing reasonably predictable behavior across a fairly flexible range of applications and operating conditions.
Complementarily, there are operating conditions that are known to be unworkable, such as using GFP in an anaerobic environment.
Practitioners who have applied these tools have acquired a great deal of pragmatic knowledge about their range of flexibility and the conditions under which their behavior can be predicted.
Some of this accumulated knowledge has found its way into scientific publications, but much of it is still communicated only through informal channels and by word of mouth. 

Functional Synthetic Biology proposes that we should begin capturing such knowledge in device specifications.
Prior work on predictive modeling (e.g.,~\cite{Salis09,davidsohn2015accurate,nielsen2016genetic}) and  reproducibility (e.g.,~\cite{beal2018quantification,beal2020robust,pine2016evaluation}) can provide initial information for some common devices.
Similar information can be captured for other devices by collecting it from experts and the literature or by conducting similar studies.
Such knowledge can then be applied, if desired, in optimization processes, particularly multi-objective optimization that can assign value to flexibility (e.g.,~\cite{boada2016multi}).
Flexibility and predictability of devices may also be improved by applying methods for insulating devices from context (e.g.,~\cite{RiboJ12, mutalik2013BCDs, carr2017reducing}).

\begin{figure*}
\centering
\includegraphics[width=0.9\textwidth]{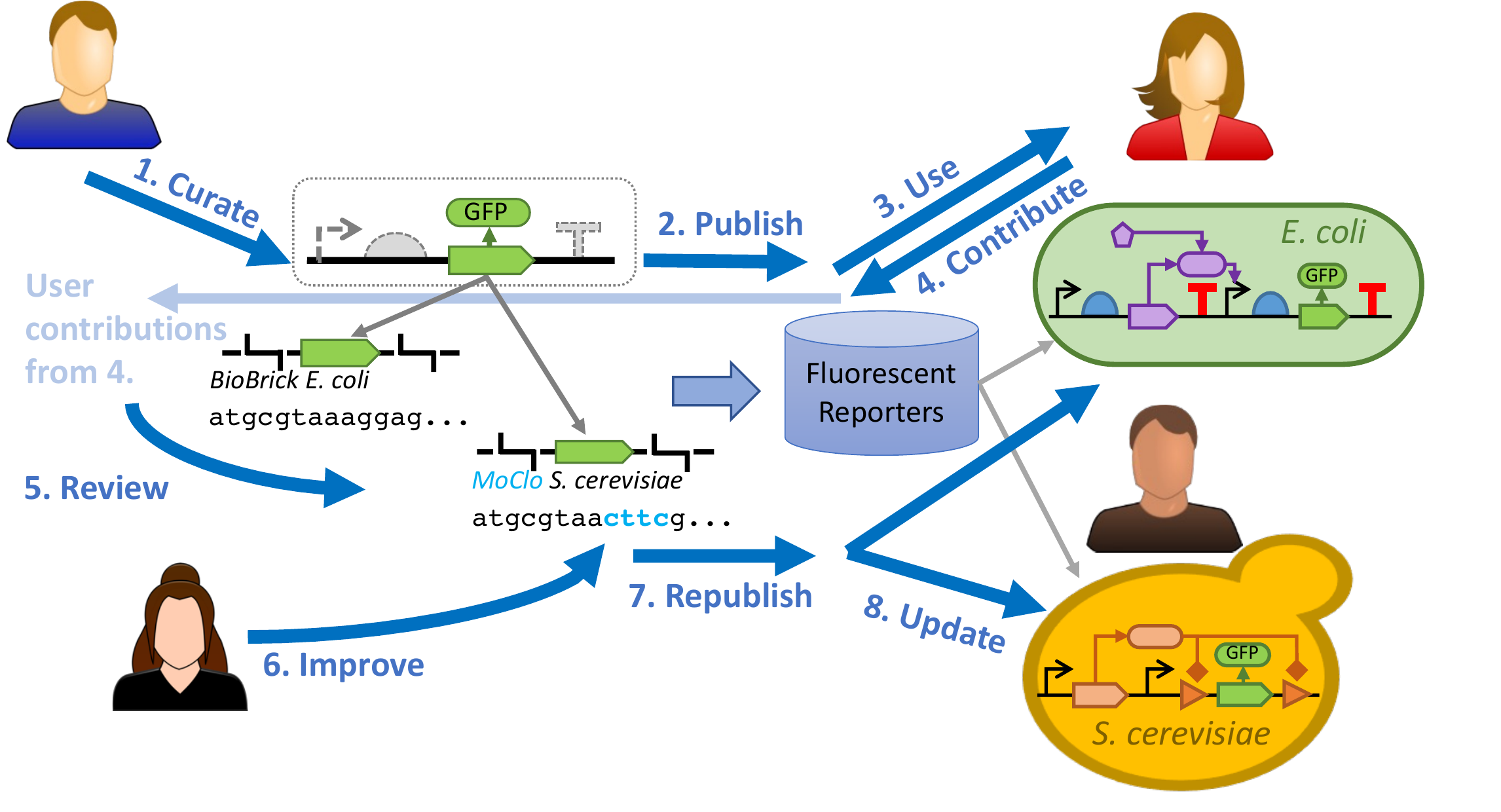}
\caption{Decoupling function and sequence enables the development of a collaborative ecosystem for distributing, using, and improving biological devices.
In this vision, (1) an expert curates a device and a set of parts that can implement the device, (2) then publishes these in a collection where they can be discovered by other synthetic biologists, (3) who put the device to use in various contexts. 
Those practitioners may (4) contribute back ``patches'' to improve the device, e.g., improved characterization or context tolerance information, (5) which are then reviewed by either the original expert or others helping to maintain the collection.
The experts may also (6) improve the collection in other ways, e.g., by improving the design of the parts that implement the device. All of these improvements are then (7) made available when the collection is republished as a new version, and (8) the device users receive the benefits of these improvements by updating to the newest version.}
\label{f:workflow}
\end{figure*}

In sum, the focus on the functionality of a device rather than the sequence of a part reinforces the need for better documentation that can usefully describe the behavior, context, and constraints of a genetic object in the same way that other engineering disciplines seek to describe the specifications of a component.
It also drives a need for characterization focused on improving the understanding of composability, flexibility, predictability, and robustness. 
Explicitly studying and recording such information in a tool-friendly manner will allow the information to be shared and redistributed more broadly, and will allow practitioners to more readily benefit from knowledge and advances produced by others in the field.

\section{Collaborative Reduction of Technical Risk}

Decoupling functional specifications and part sequences also allows new approaches to collaboration and information sharing that can reduce the risk of failure in engineering synthetic biology systems.
Here the key idea is that changes in parts do not necessarily entail changes in devices or vice versa.

When a part is improved, the new version that results may still conform to the same device specifications that the old version satisfied, and thus be able to be substituted as an implementation for that device.
For example, a GFP part might be codon optimized to produce the protein more efficiently, or have a restriction site eliminated to make it compatible with a wider range of assembly methods.
If the new version of the part still satisfies the Green Fluorescence Reporter device specification, then it can be safely predicted that systems using the old version can be upgraded to use the new version.
Likewise, a device may be improved with models that better predict its interactions with other devices, or with better documentation of its expected operational range, while having no impact on the sequence of parts that implement the device.

This decoupling offers a new means of capturing expert knowledge, in the form of curated collections of devices and the recommended parts to implement them.
For example, a typical synthetic biologist should never need to consider the 160 different variations of GFP in FPbase, let alone other green proteins like ZsGreen or mNeonGreen.
Instead, they should be able to determine their intended operating range (e.g., {\it E. coli} DH5$\alpha$ in M9 media at 30 $^{\circ}$C to 37 $^{\circ}$C), select a Green Fluorescence Reporter device that operates in that range, and then implement it with any of the parts that are currently recommended by experts as a good implementation for that device.

If better parts become available or a problem is detected with one of the current parts, then all that needs to be changed is the recommendation.
As long as the functional characteristics of the newly recommended part can be assessed to remain within the range of predictability that has been established for the Green Fluorescence Reporter device, any system using the device should be able to be safely updated to use the new part.
Indeed, such an upgrade recommendation can already be found in the scientific literature for red fluorescence, when the developers of mCherry suggested phasing out its predecessor, mRFP1, given mCherry's improved ``higher extinction coefficient \dots, tolerance of N-terminal fusions and photostability.''~\cite{shaner2004improved}

Assessing the potential impact from a change to a part or device, however, is often not clear-cut or straightforward.
If a model is not sufficiently accurate, there is a risk that changes assessed as safe will instead produce unwanted side-effects or even result in a system-wide degradation. 
This risk from adopting a recommendation, however, must be balanced against the risk and costs associated with ignoring a recommendation from experts who are likely to be more familiar with the specialized matter at hand and better able to sort through the myriad possible implementations of the device.
Ultimately, then, this is a question of building trust around changes to complex systems with implications that are difficult to predict.

The software engineering community has faced a very similar problem of managing technical risk and building trust around changes with difficult-to-predict systems implications.
That community has addressed its analogous challenge with a now-mature ecosystem of tools and collaborative processes (often collected under the name ``agile software development'') for managing the development of complex systems.

Foundational to these processes are distributed version control systems (e.g., git) that afford communities of experts a convenient way to organize, share, and maintain packages of information. 
These are already used in the bioinformatic community to manage knowledge collections such as the Systems Biology Ontology (SBO), Sequence Ontology (SO), and Gene Ontology (GO).

\begin{figure*}
\centering
\includegraphics[width=\textwidth]{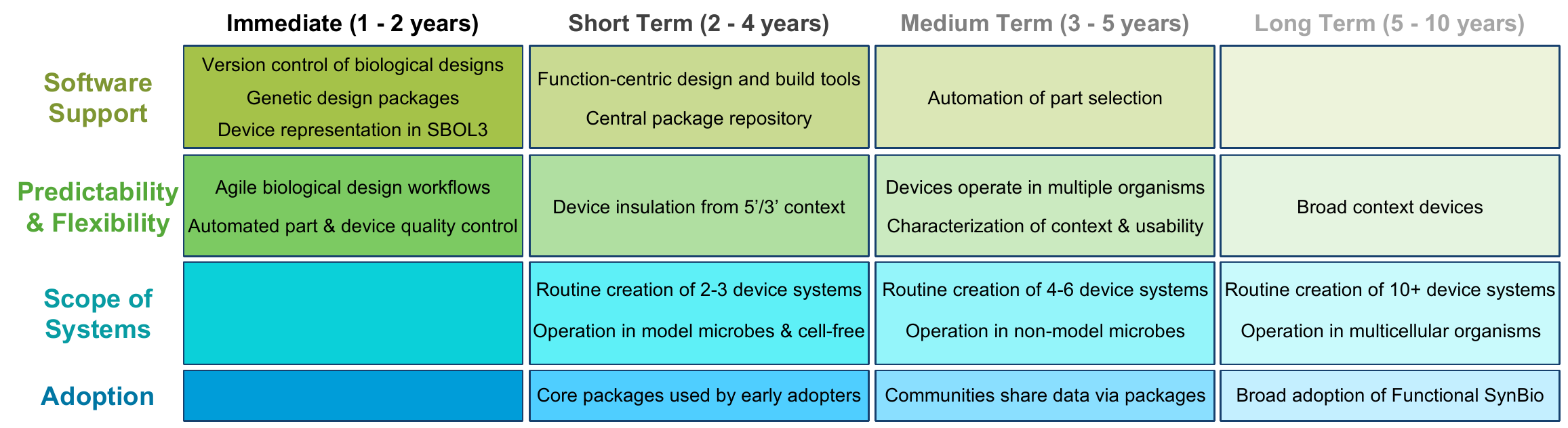}
\caption{Suggested roadmap of goals, from immediate to long-term, for achieving Functional Synthetic Biology.}
\label{t:roadmap}
\end{figure*}

Functional Synthetic Biology can take advantage of these same mechanisms to curate collections of devices, collaboratively maintain them, and reduce technical risk related to updating these collections.
Figure~\ref{f:workflow} illustrates the type of interactions that can be enabled.
In this vision, an expert curates a device such as a Green Fluorescence Reporter, along with options for implementing the device using one of several parts containing a coding sequence for GFP.
This device is aggregated with others into a fluorescent reporters collection, which can be published as a package in a catalog where it can be shared with other synthetic biologists.

In the course of applying the device in various contexts, synthetic biologists will undoubtedly identify ways to improve the collection.
Those improvements might include expanding or sharpening characterization, augmenting context tolerance information, clarifying documentation, or enhancing the design of the parts that implement the device.

Distributed version control can make it easy for users to contribute their improvements back as suggested changes to the collection.
Agile software tooling can support the curation of such contributions, making it simple to the maintainers of a collection to review and discuss a contribution, request changes where needed, and apply automation-assisted validation checks for quality control.
Once incorporated, improvements can be taken up in a new version of the collection that is made available to the community when the collection is republished.
Users can then take advantage of the benefits these curated improvements provide simply by updating their copy of the collection to the latest version.

When implemented well, the transparency and checking in such processes can help build trust within a community of users and maintainers.
The mechanisms of distributed version control also help to sustain an open marketplace for competition between packages of information and their attendant processes and maintainers.
Just as in the software world, Functional Synthetic Biology packages will compete not just on the basis of technical efficacy, but also their trustworthiness, reliability, ease of use, and responsiveness to user needs.

\section{A Roadmap to Functional Synthetic Biology}


In this section, we propose a multi-year roadmap for realizing the vision for Functional Synthetic Biology set out above. 
This roadmap foregrounds the vision's most consequential aspects and specifies the timeframes within which we predict that these aspects can reasonably be achieved (Figure~\ref{t:roadmap}).

Some aspects of the vision that the roadmap accounts for are well-defined and can be achieved with technologies and techniques that are available today. They simply need to be adapted and applied. 
For example, an integrated representation of the part and device hierarchies from Figure~\ref{f:function} can already be implemented using the SBOL~2~\cite{roehner2016sharing} or SBOL~3~\cite{mclaughlin2020synthetic} standards.
Alternatively, devices could also be represented with a modeling language such as SBML~\cite{SBML}, then associated with parts represented in GenBank or GFF.
In either case, a modern version control system such as git can be employed to manage and disseminate the genetic design files generated. 
Another advantage of such a version control system
is that it supports mature agile software workflows (e.g., trunk-based development) and tooling (e.g., continuous integration) that foster collaboration and enable quality control automation.

Short term targets for development include basic parts such as constitutive promoters, terminators, sensors, and reporters, as well as regulatory insulation (e.g.,~\cite{RiboJ12, mutalik2013BCDs, carr2017reducing}) and families of transcriptional computing devices (e.g.,~\cite{endy-pnas-2012, kiani2014crispr, nielsen2016genetic, gander2017digital, chen2020genetic}).
Developing a collection of design packages covering these elements, along with function-centric design and build tools, should enable routine engineering of 2-3 device sense/compute/actuate circuits in model microbes and cell-free systems.
The core packages that result from these initial engineering initiatives can be made available through a central repository serving as a rendezvous point for early adopters to discover and make use of these packages.

As this Functional Synthetic Biology ecosystem matures and expands, practitioners will undoubtedly enhance the performance, flexibility, and reliability of the parts, will add new packages focused on their own areas of interest, and will expand the functionality of existing packages by making incremental upgrades to them. 
The complexity of systems that can be routinely engineered will increase, as will levels of automation.

The ability to support more effective sharing of data will then facilitate better understanding and characterization of operational context, expansion into non-model microbes, and the creation of more context-agnostic devices that can operate effectively in multiple types of organisms.
Over the longer term, we anticipate a widespread adoption of Functional Synthetic Biology, driven by the standardization of complex biological system engineering, the development of extremely flexible devices, and extension even to effective operation {\it in vivo} in complex multicellular organisms.

Following this roadmap will also require dealing with non-technical challenges regarding incentives.
For example, an initial investment in time and resources is needed from experienced practitioners before the community can benefit, and the people investing will not necessarily be the ones who most benefit.
Given the struggle to even obtain DNA sequence information from scientific publications~\cite{peccoud2011essential}, we can expect that current academic incentives will not be sufficient on their own to motivate widespread investment in the curation and publication of functional information.
There will also be questions around governance and degree of centralization for the evolving collection of functional packages.
Finally, there is an important collective action problem to be solved if the community wishes resources to be available under free and open licenses, as opposed to needing to pay for access to this information.

\section{Conclusions}

Many practitioners of synthetic biology recognize the impact that standardizing systems-level engineering could have on the development of biological systems and are eager to exploit the potential it has to democratize access to complex biotechnology and effect transformative change in a broad range of sectors and application areas such as healthcare, manufacturing, energy, agriculture, and environmental sustainability.

To date, however, realizing that potential has proven elusive, in large part because the information required for effectively engineering with biological parts and devices has been inaccessible, insufficient or incomplete. 
The Functional Synthetic Biology framework presented here lays the groundwork for a shift in  orientation from the focus on sequences that defines the field today to a focus on functionality that will transform the way synthetic biology is practiced in the future.  
This represents a critical step forward along the path to achieving systems level engineering, with improvements in data sharing leading to increases in flexibility and predictability, which in turn open up opportunities for enhanced collaboration and dissemination.
It is our hope that this paper will encourage others to build on the framework presented, to join us on our journey to transform the practice of synthetic biology using the roadmap we have laid out as a guide, and to help mobilize the resources that will be required to bring that journey to a successful conclusion.

\section{Competing interests}
No competing interest is declared.

\section{Material Availability Statement}
Not applicable.

\section{Data Availability Statement}
Not applicable.

\section{Author contributions statement}
All authors contributed to conceptualization, writing the original draft, and review and editing.

\section{Funding}

This work was supported in part by the following funding sources:
J.Be. was supported by Air Force Research Laboratory (AFRL) and DARPA contract FA8750-17-C-0184.
J.Bo. was supported by a Horizon Postdoctoral Fellowship from Concordia University.
M.H. was supported by funding from the ERC Advanced Grant LoopingDNA (no. 883684).
N.S. was supported by funding from the BBSRC under award BB/M011178/1.
G.V. was supported by a scholarship from the School of Computing, Newcastle University.
A.V. was funded by Grant MINECO/AEI, EU DPI2017-82896-C2-1-R and MCIN/AEI/10.13039/501100011033 grant number PID2020-117271RB-C21.

This document does not contain technology or technical data controlled under either the U.S. International Traffic in Arms Regulations or the U.S. Export Administration Regulations.
Views, opinions, and/or findings expressed are those of the author(s) and should not be interpreted as representing the official views or policies of the Department of Defense or the U.S. Government.


\end{document}